\documentstyle[epsf,graphics,epsfig]{mn}

\newcommand{\asec}{^{\prime\prime}}

\title[AGN Cluster Environments]{The cluster environments of powerful radio-loud and radio-quiet AGN}
\author[R.J. McLure]
{R.J. McLure$^{1}$ \&\ J.S. Dunlop$^{2}$\\
 $^{1}$Nuclear and Astrophysics Laboratory, University of Oxford, Keble Road, Oxford, OX1 3RH\\
 $^{2}$Institute for Astronomy, University of Edinburgh, Blackford Hill, Edinburgh, 3H9 3HJ}
\date{Submitted for publication in MNRAS}
\begin{document}
\maketitle
 
\begin{abstract}
The spatial clustering amplitude ($B_{gq}$) is determined for a sample 
of 44 powerful AGN at $z\simeq0.2$. No significant difference is detected
in the richness of the cluster environments of the radio-loud and
radio-quiet sub-samples, both of which typically inhabit environments
as rich as Abell Class $\simeq 0$. Comparison with radio 
luminosity-matched samples from Hill \&\ Lilly (1991) and Wold et al. (2000a) 
suggests that there is no epoch-dependent change in environment
richness out to at least $z\ge0.5$ for either radio galaxies or radio
quasars. Comparison with the APM cluster survey shows that, contrary
to current folklore, powerful AGN
do not avoid rich clusters, but rather display a spread in cluster
environment which is perfectly consistent with being drawn at random
from the massive elliptical population. Finally, we argue that
virtually all Abell class $\simeq0$ clusters contained an active
galaxy during the epoch of peak quasar activity at $z\sim2.5$.
\end{abstract}
 
 
\section{Introduction}
 
The cluster environments of active galactic nuclei (AGN) provide
important information for improving our
understanding of several 
aspects of the AGN phenomenon. Over the last fifteen years a
considerable amount of effort has been
invested in studying the environments of several different types of AGN, 
including : radio-quiet and radio-loud quasars (eg. Yee \&\ Green 
1984, 1987, Smith, Boyle \&\ Maddox 1995, 2000, Hall \&\ Green 1998), 
radio galaxies  
(Prestage \&\ Peacock 1988, 1989, Hill \&\ Lilly 1991), BLLacs (Wurtz et al. 1997) 
and Seyfert galaxies (de Robertis \&\ Yee 1998). In addition to its
potential for shedding light on the
processes by which dormant black holes are triggered into AGN, the study of
clustering environments is an invaluable test of the viability of proposed
unification schemes (eg. Wurtz et al. 1997) and models of quasar 
evolution (eg. Ellingson, Green \&\ Yee 1991). In this paper we
use {\sc hst} images to study the 
immediate environments of AGN drawn from well-matched
samples of powerful radio galaxies (RG), 
radio-loud quasars (RLQ) and radio-quiet quasars (RQQ) at $z=0.2$ to
explore what constraints can be placed on
the origin of radio loudness, the viability of radio loud unification,
and the physical origin of the dramatic evolution of the quasar population
between z = 2.0 and the present day.

 Several authors have reported a difference in the richness of
 clustering around radio-loud and radio-quiet quasars (eg. Yee \&\
 Green 1984, 1987, Ellingson, Green \&\ Yee 1991). These studies
 have shown that, with substantial overlap, low to moderate redshift
 radio-loud quasars are found in Abell 0/1 clusters, while radio-quiet
 quasars rarely inhabit clusters as rich as Abell 0. This apparent
 change in quasar environment with radio-power was at least consistent with
 the traditional picture in which RLQs have elliptical host
 galaxies, and are therefore preferentially found in clusters, while
 RQQs inhabit the poorer environments associated with lower
 luminosity disc-dominated Seyfert galaxies.

However, the results of our own {\sc hst} host galaxy imaging
programme (McLure et al. 1999, Dunlop et al. 2000), together with
other recent {\sc hst} and ground-based studies (eg. Hooper et
al. 1997, Boyce et al. 1998, Bahcall et al. 1997, Schade et al. 2000)
show that, at the very least, a significant fraction of the RQQ
population are located in elliptical host galaxies. In fact, the
results of our own study lead to the conclusion that all objects which
could be classified as {\it true} quasars (i.e. M$_{V}<-23$) will have
bulge-dominated host galaxies. In Section \ref{rqq/rlq} we therefore
examine the question of whether
our target RQQs and RLQs, drawn from samples well matched in terms of AGN
luminosity, are actually occupying significantly different cluster environments.

The other question investigated in this paper is that of the evolution
of the cluster environments of powerful, radio-loud AGN. With respect to radio
galaxies, it was shown by Prestage \&\ Peacock (1988, 1989) that at
low $z$ the environments of powerful {\sc frii} sources are poorer
than their {\sc fri} counterparts, although the substantial overlap
between the two populations led Prestage \&\ Peacock to conclude that
environment did not solely determine radio luminosity. In contrast, at
higher redshifts, $z\simeq 0.5$, both Yates, Miller \&\ Peacock (1989)
and Hill \&\ Lilly (1991) found that radio galaxies inhabited richer
clusters than at low redshift, leading Hill \&\ Lilly to conclude that the
cluster environments of radio galaxies were epoch-dependent. Similar
conclusions were reached by Ellingson, Green \&\ Yee (1991) from
their study of the environments of RLQs. Ellingson et al. found that the
environments of RLQs also become richer at $z>0.6$, although the
existence of this epoch-dependence has recently been questioned by
Wold et al. (2000a).

In contrast to what might be expected from the results outlined above,
the raw {\sc hst} images from our sample of RGs and RLQs (Dunlop et
al. 2000) show many of the objects to be apparently residing in
clusters of moderate richness, an impression which is confirmed by the
analysis presented in Section \ref{ressec}. Motivated by this, in
Sections \ref{hillcomp} \&\ \ref{woldcomp} we
present a comparison between the cluster results for our RG and RLQ
samples with those for optical and radio luminosity-matched sub-samples
of the objects studied by Hill \&\ Lilly (1991) and Wold et
al. (2000a), in order to critically re-examine the evidence that RGs and
RLQs inhabit environments which are a function of cosmic
epoch. In Section 6 we investigate whether there is any evidence in
our data for a link between AGN properties and the large-scale
environments of their hosts. Finally, in Section
\ref{apm} we compare the distribution of cluster richness determined
for our sample with that of the APM cluster survey (Dalton et
al. 1997, Croft et al. 1997) in light of the constraint imposed by our 
host galaxy study
that only galaxies with M$_{R}\leq-23$ are capable of producing
powerful AGN.  Unless otherwise specified all cosmological calculations 
performed in this paper assume a cosmology of H$_{0}=50$, q$_{0}=0.5$ 
and $\Lambda=0$.
 
\section{The Sample}
 
 The low-redshift sample studied in this paper comprises 44 objects and is a
 combination of that investigated in our {\sc hst} study of $z\simeq0.2$ AGN
 host galaxies (McLure et al. 1999, Dunlop et al. 2000), with data of
 similar quality from the {\sc hst} archive. The sample investigated in
 our host-galaxy programme consisted of 33 objects, divided into
 three sub-samples of 10 radio galaxies (RG), 10 radio-loud quasars
 (RLQ) and 13 radio-quiet quasars (RQQ). The unique feature of this
 sample is that the two quasar sub-samples were originally chosen to be
 matched in terms of their distribution in the optical
 luminosity-redshift plane (M$_{V}-z$) (Dunlop et al. 1993), with the 
 RG and RLQ sub-samples matched in both the radio power-redshift plane
 (P$_{5GHz}-z$), and the radio power-spectral index plane
 (P$_{5GHz}-\alpha$) (Taylor et al. 1996). Although the selection
 criteria used in choosing
 these samples were designed to determine the role played by host
 galaxies in the radio-loudness dichotomy and radio-loud unification, 
 they are equally valid for investigating the nature of their respective 
 cluster environments. One of the objects from this sample, the RQQ
 1549+203, has however been excluded from this analysis due to the
 known presence of a foreground cluster (Stocke et al. 1983). In order
 to increase the numbers of objects
 studied we have also included in our sample 12 quasars from the
{\sc hst} study of Bahcall et al. (1997) which were not included in our
 host galaxy study. Although image saturation and
 emission line contamination make the Bahcall et al. data flawed with
 respect to detailed host galaxy analysis, these problems are not a
 concern for an environmental study. The data for these objects 
(3 RLQs \&\ 9 RQQs) is of practically identical depth to our own and, 
with the exception of  3C273, falls in the same region of the M$_{V}-z$ plane.

\subsection{Observations and data reduction}
 
 The observations for all 44 objects considered in this paper were
 taken with the {\sc hst} Wide-Field and Planetary Camera 2 (WFPC2). The
 observations for the 33 objects from our $z=0.2$ host galaxy study
 utilised the F675W filter, which closely approximates the
 standard Cousins $R$-band (Holtzman et al. 1995), while the data for the
 objects from the Bahcall et al. sample were taken with the F606W
 (Wide $V$) filter. Both data sets were imaged on the wide-field chips
 of WFPC2 (WF2 for the F675W data and WF3 for the F606W data) and have a
 plate-scale of $0.1\asec$/pix. 
 
 The basic data reduction (bias removal \&\ flat-fielding)
 was performed by the {\sc hst} pipeline. Subsequently, the individual
 exposures of each source ($3\times 600$s for the F675W data and
 typically 1100s+600s for the F606W data) were then combined and
 cleaned of cosmic rays using standard {\sc iraf} tasks. The
 signal-to-noise levels in the final deep images correspond to a
 Cousins $R$-band $1\sigma$ sensitivity limit of $\mu_{R}\simeq26.7$
 mag arcsec$^{-2}$, where the 
 conversion from F606W to Cousins $R$-band for the Bahcall et al. 
data assumes a typical colour for a $z=0.2$ E/Sab galaxy of $R_{c}=606-0.3$ 
 (Fukugita et al. 1995).
 
\section{Determining Cluster Richness}
\label{spam}
 The method adopted in this study for quantifying the richness of the
 AGN  environments was to determine their respective spatial
 clustering amplitudes ($B_{gq}$) (Longair \&\ Seldner 1979). This is a
 standard technique which, although originally designed to investigate 
clustering around radio
 galaxies, has subsequently been successfully applied to the
 environments of a wide variety of active and inactive galaxies 
(eg. Yee \&\ L\'{o}pez-Cruz 1999, Ellingson, Yee \&\ Green 1991,
 Prestage \&\ Peacock 1988). A
 detailed description of the derivation of $B_{gq}$ is given in 
Longair \&\ Seldner (1979) and consequently only a brief outline is 
provided here. The first stage in the calculation is the determination 
of the angular correlation function, defined as :
\begin{equation}
 n\left( \theta \right)\delta\Omega=N_{g} \left[ 1+w \left( \theta \right)
  \right] \delta \Omega
\end{equation}
\noindent
 where
\begin{equation}
 w \left( \theta \right)=A_{gq} \theta^{1-\gamma}
\label{angcor}
\end{equation}
\noindent
 and A$_{gq}$ quantifies the excess in the number of galaxies in the
 vicinity of the source as compared to the expected background
 contribution N$_{g}$. Provided that $\theta \ll 1$ the value of A$_{gq}$
 can be directly calculated from the data using the expression:
\begin{equation}
 A_{gq}=\left[ \frac{N_{t}}{N_{b}} -1 \right]\left(\frac{3-\gamma}{2} \right)\theta^{\gamma-1}
\end{equation}
\noindent
 where $N_{t}$ is the total number of galaxies counted within a
 radius of $\theta$ radians from the target (excluding the target itself), and
 $N_{b}$ is the expected number of 
background galaxies within the same radius. In order to directly compare
 the clustering around objects which cover a range of redshifts it is
 then necessary to de-project this angular correlation into its spatial
 equivalent which is defined by:
\begin{equation}
 n\left(r\right)\delta V=\rho_{g}\left[1+\epsilon\left(r\right)\right]\delta V
\end{equation}
\noindent
 where
\begin{equation}
\epsilon\left(r\right)=B_{gq}r^{-\gamma}
\end{equation}
 and $B_{gq}$ is the desired spatial clustering amplitude. By invoking the
 simplifying assumption that the clustering of galaxies is spherically
 symmetric around the central object it can be shown (Longair \&\
 Seldner 1979) that the spatial and angular clustering amplitudes are
 related by the expression:
\begin{equation}
 B_{gq}=\frac{A_{gq} N_{g}}{ I_{\gamma}\phi\left(z\right) }
\left[\frac{D}{1+z}\right]^{\gamma-3}
\label{bequn}
\end{equation}
\noindent
 where $D$ is the effective angular diameter distance to the target and
 $\phi\left(z\right)$  is the integrated field-galaxy luminosity
 function at the redshift of the target.
 The quantity $I_{\gamma}$ is a constant which has a value of $3.78$
 for the canonical field-galaxy value of $\gamma=1.77$ (Groth \&\ Peebles 1977). 
 
 For a galaxy appearing on the same WF CCD as one of the AGN in this
sample to be counted as a possible cluster member it had to satisfy
two criteria. Firstly, its projected distance from the central object
had to be less than the counting radius; defined as the distance of
the central object from the nearest edge of the CCD. At the sample median
redshift of z=0.2 this radius corresponds to a projected metric radius 
of typically 180 kpc. Although this is undoubtedly a small counting
radius compared to the usual 500 kpc or 1 Mpc counting radii adopted
in most cluster studies, provided that the clustering of galaxies
around the AGN does have a slope of $\gamma=1.77$, the restriction of
only having information of the central regions of the clusters should 
not prevent a reliable determination of the enhancement of associated 
galaxies relative to the field population. Indeed, as will be
discussed in Section \ref{corrslope}, there is good evidence that the 
clustering around the AGN studied here does follow the expected slope 
of $\gamma=1.77$, at least on average. Furthermore, in their recent 
study of low-$z$ Abell clusters, Yee \&\ L\'{o}pez-Cruz (1999) found that
reducing the counting radius from 1 Mpc to 200 kpc only altered the 
spatial clustering amplitude determination at the 10 percent level. 
 
 The second selection criterion was that galaxies had to lie in the
magnitude interval $m_{\star}-1\rightarrow m_{\star}+2$, where m$_{\star}$ is 
the apparent magnitude corresponding to M$_{\star}$ at the AGN 
redshift. This magnitude interval is a compromise which is designed to include 
those cluster galaxies which contain the majority of the 
cluster mass. The faint limit of m$_{\star}+2$ should probe deep enough 
into the cluster luminosity function to be sensitive to any
 enhancements in galaxy density, without risking the possibility 
of missing clusters due to the background galaxy counts rising 
more rapidly than the faint-end of the cluster luminosity function 
at $m>m_{\star}+2$. The bright limit of m$_{\star}-1$ is set to 
avoid problems associated with small number statistics, although 
relaxation of this limit to m$_{\star}-3$ has a negligible effect
 on the results presented in Section \ref{ressec}. An additional 
advantage with this choice of magnitude interval is that, due to
 the host galaxies of the AGN studied in our {\sc hst} imaging 
programme having average luminosities one magnitude brighter 
than M$_{\star}$ (Dunlop et al. 2000), this interval typically 
corresponds to the m$_{g} \rightarrow$m$_{g}+3$ range adopted 
for the Abell-type calculation of cluster richness around $z\simeq0.5$
 radio galaxies by Hill \&\ Lilly (1991). The comparison presented 
in Section \ref{hillcomp} of the clustering around the $z\simeq0.2$
 radio galaxies studied here, with the results obtained by 
Hill \&\ Lilly (1991), can consequently be performed in a more
 transparent manner.
 
\subsection{Galaxy counts and the luminosity function}
\begin{table}
\begin{center}
\begin{tabular}{cccc}
\hline
\hline
 z&$\phi$/Mpc$^{-3}$&M$_{R}^{\star}$&$\alpha$\\
\hline
 $0.00\rightarrow0.20$ &0.0023&$-22.20$&$1.00$\\
 $0.20\rightarrow0.50$ &0.0034&$-22.32$&$1.03$\\
 $0.50\rightarrow0.75$ &0.0078&$-22.11$&$0.50$\\
 $0.75\rightarrow2.00$ &0.0068&$-22.52$&$1.28$\\
\hline
\hline
\end{tabular}
\caption{The Schechter luminosity function parameters used to predict
 the background galaxy counts. Column 1 lists the four redshift bins
 adopted. The parameters for the lowest redshift bin have been taken
 from Yee \&\ L\'{o}pez-Cruz (1999). The parameters in the other three
 redshift bins have been taken from Lilly et al. (1995) with a
 conversion of M$_{\star}$  deduced assuming typical colours of $B-B_{AB}=0.17$
 (Metcalfe et al. 1991) and $R-B=1.45$ (Fukugita et al. 1995).}
\label{lumfn}
\end{center}
\end{table}
\begin{figure}
\centerline{\epsfig{file=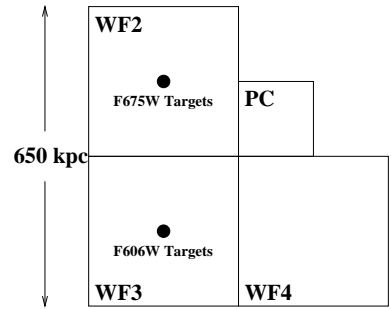,width=5.0cm,angle=0,clip=}}
\caption{A schematic of the Wide Field and Planetary Camera 2 showing
 the spatial coverage at the median redshift of $z=0.2$. The 32
 objects imaged during the AGN host galaxy study of McLure et
 al. (1999) and Dunlop et al. (2000) are centred on WF2. The 12 objects
 included from the sample of Bahcall et al. (1997) were imaged on
 WF3. The extra separation of WF2 and WF4 allowed the distribution of
 background galaxies and the slope of the correlation function to be
 investigated (Section \ref{spam}).}
\label{wfscheme}
\end{figure}
 
\begin{figure}
\centerline{\epsfig{file=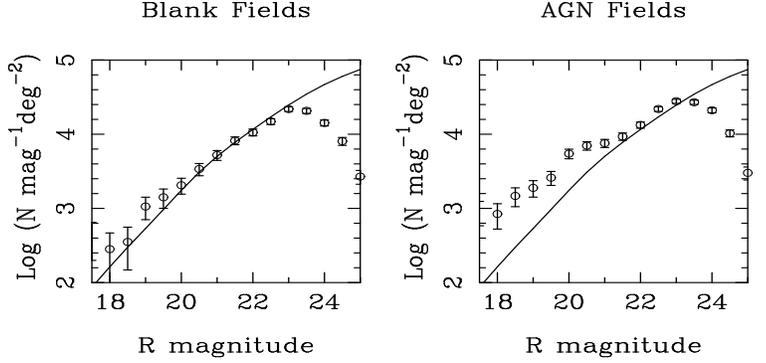,width=10.0cm,angle=0,clip=}}
\caption{The left-hand panel shows the background galaxy number counts 
from the 32 WF4
 observations from McLure et al. (1999) and Dunlop et al. (2000), at an
 average separation from the target AGN of $\simeq 500$ kpc. The solid
 line is the predicted number counts from integrating the galaxy luminosity
 function along the line of sight (see text for discussion). The right-hand
panel shows the  galaxy counts from the target WF chips of all 44
objects at a separation of $\le 200$ kpc. A clear excess of galaxies
is obvious at $R<23$.} 
\label{counts}
\end{figure}

 A crucial element in the calculation of the spatial clustering
 amplitude is the form of the luminosity function used in the
 normalization of equation \ref{bequn}. Given that the derivation of
 equation \ref{bequn} is dependent on the assumption that the
 observed background galaxy counts can be predicted from integrating the
 galaxy luminosity function along the line of sight, it is essential
 that the luminosity function chosen should at least be consistent with
 the background counts. In order to ensure this we have adopted a
 similar approach to that employed by Wold et al. (2000a), by considering
 the form of the luminosity function in four redshift bins
 z=(0.0, 0.2), (0.2, 0.5), (0.5, 0.75) \&\ (0.75, 2.0). The Schechter
 function parameters of the four luminosity functions are listed in
 Table \ref{lumfn}. In the lowest redshift bin we have adopted the
 luminosity function determined by Yee \&\ L\'{o}pez-Cruz (1999) from their 
 $R$-band study of Abell clusters in the redshift range
 $0.02<z<0.18$. 
The characteristic magnitude and
 faint-end slope of this Schechter function are nearly identical to
 those determined for the field-galaxy population by Loveday et al. (1992),
  but with a normalization some $30\%$ higher. Due to the fact that 
 the Loveday et al. determination from the Stromlo-APM survey is
 dominated by objects 
with $z<0.1$, it was felt that the Yee \&\ L\'{o}pez-Cruz luminosity
 function was a fairer representation of the $z\simeq0.2$ galaxy
 population. The adoption of the higher normalization also has the
 advantage that we are correspondingly less likely to be overestimating
 the spatial clustering amplitudes. In the three
 highest redshift bins the parameters are taken from Lilly et al. (1995), 
 with the appropriate conversion of the
 characteristic magnitudes from the M$_{AB}$ system to the Cousins
 $R$-band (see Table \ref{lumfn}). The predicted number counts in a 
 particular magnitude range $(m_{1},m_{2})$ were then calculated by
 integrating the following function:
\begin{equation}
\int_{z=0}^{z=2}\int_{m_{1}}^{m_{2}}\phi(m,z)\delta m
\left(\frac{\delta V}{\delta z}\right)\delta z
\end{equation}
\noindent
 where the limiting redshift of $z=2$ was chosen since, in the
 magnitude range investigated here ($R<24$), the contribution from
 background galaxies at $z\ge2$ is negligible. A comparison between the
 predicted and measured galaxy counts for both the target (WF2+WF3) and control
 CCDs (WF4) is shown in Fig \ref{counts}. It can be seen that the predicted
 background galaxy counts (left panel) agree well with those measured
 in the
 magnitude range ($21<R<23$), where the control chips should be relatively
 unaffected by contamination from cluster members. However,
 in the magnitude range $18.5<R<20.5$ there is an excess of detected
 background galaxies over that predicted, as expected given that 
 at our median redshift of $z=0.2$ this range corresponds to
 $M_{\star}\rightarrow M_{\star}+2$. The predicted background
 counts are in good agreement with those determined by both Yee \&\
 L\'{o}pez-Cruz (1999) and Metcalfe et al. (1991) and were subsequently used to
 remove the background contribution in the calculations of the spatial
 clustering amplitude. The comparison between the
 predicted background counts and the counts from the AGN target CCDs
 shows qualitatively that inside a radius of $\simeq200$ kpc we are clearly detecting a large excess of cluster galaxies. 
\subsection{The slope of the correlation function}
\label{corrslope}
\begin{figure}
\centerline{\epsfig{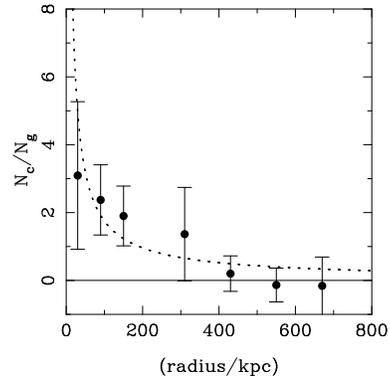}}
\caption{Radial profile showing the average ratio of excess cluster galaxies
 (N$_{c}$) to background galaxies (N$_{g}$). The width of the bins changes from
60 kpc to 120 kpc at a radius of 220 kpc in order to maintain comparable 
signal-to-noise. Also shown is the best-fit power-law correlation 
function (dotted line). The best fit value of $\gamma$ is 
$1.88^{+0.10}_{-0.09}$ where the errors refer to the 
$\Delta\chi^{2}-\Delta\chi^{2}_{min}=1$ confidence limit.}
\label{angleprof}
\end{figure}

 The usual procedure when determining the richness of a cluster
 environment via its spatial clustering amplitude is to assume that the
 slope of the two-point correlation function has the canonical value of
 $\gamma=1.77$ (Groth \&\ Peebles 1977). Here we investigate whether
 this is in fact a reasonable assumption for the objects in this sample.
  As a result of the relative positions of the WF2 and WF4 CCDs on
 the WFPC2 (see Fig \ref{wfscheme}) it is possible to perform galaxy
 ring counts out to a projected radius of $\simeq800$ kpc for the 32 objects 
imaged through the F675W filter, under the assumption of spherical symmetry.
  Fig \ref{angleprof} shows the average radial profile of the ratio of 
 cluster galaxies to background galaxies ($N_{c}/N_{g}$) measured in
 metric aperture bins. In order to keep the signal-to-noise level of
 the bins approximately constant their width is increased from 60 kpc
 to 120 kpc at a radius of 200 kpc. The dotted line in Fig
\ref{angleprof} shows the minimum $\chi^{2}$ fit for the angular
 correlation function, as defined in equation \ref{angcor}, which has a value
 of $\gamma=1.88^{+0.10}_{-0.09}$. It appears therefore that, at least
 on average, the form of the correlation function for the objects in
 this sample is consistent with the standard value of
 $\gamma=1.77$. Consequently, we adopt this standard value in our
 calculation of the spatial clustering amplitudes
 in order to allow easy comparison with results in the literature.
\section{Results}
\label{ressec}
\begin{figure}
\centerline{\epsfig{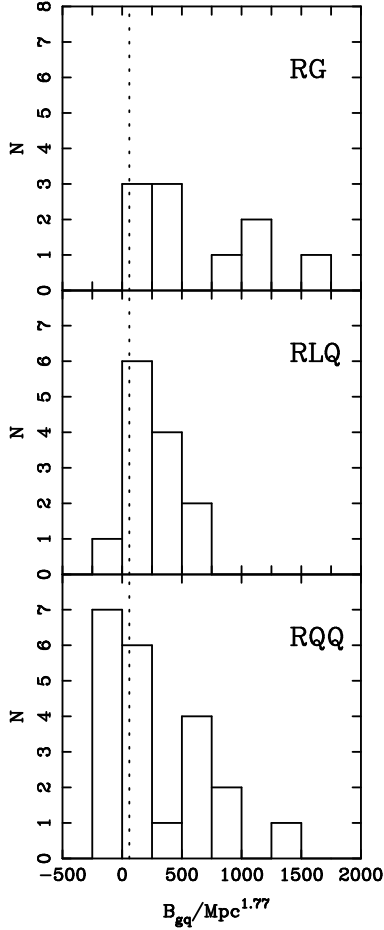}}
\caption{Histograms showing the distribution of clustering amplitudes
displayed by the three sub-samples. Also shown in the approximate level of the
clustering amplitudes of field galaxies (dotted line), here taken as B$_{gq}=60$}
\label{hist}
\end{figure}
\begin{table}
\begin{center}
\begin{tabular}{cccc}
\hline
\hline
 Sample&N&$<B_{gq}>$&Median\\
\hline
 All&44&$365\pm62$ &241\\
 RG&10 &$575\pm165$&321\\
 RLQ&13&$267\pm51$ &247\\
 RQQ&21&$326\pm94$ &209\\
\hline
\hline
\end{tabular}
\caption{Summary of the spatial clustering amplitude results.
 Column two details the number of objects in each sub-sample.}
\end{center}
\label{tab2}
\end{table}
 
\begin{figure*}
\centerline{\epsfig{file=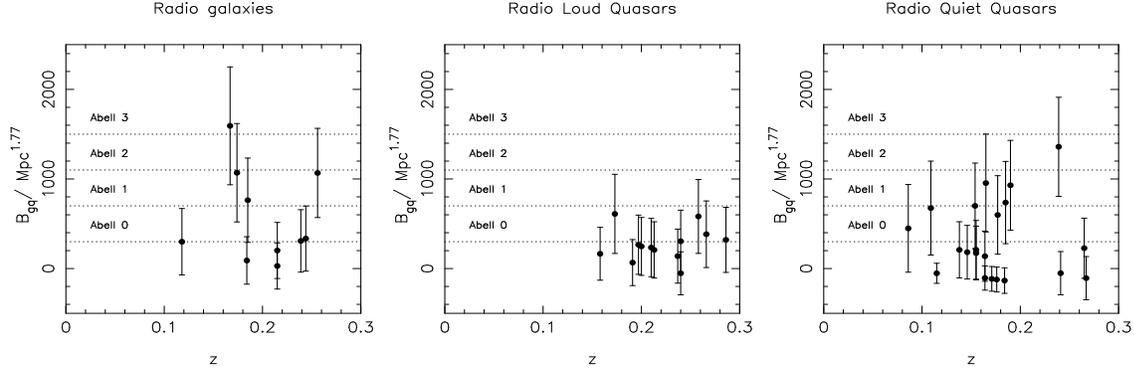,width=15.0cm,angle=0,clip=}}
\caption{The B$_{gq}-z$ distribution for the three sub-samples.  Also
shown are the approximate Abell cluster
classifications according to the linear scheme shown in Table 4.}
\label{spamfig}
\end{figure*}

The results of the $B_{gq}$ calculations for the three sub-samples are
listed in column 6 of Table \ref{bigtab} and are shown graphically in Figs
\ref{hist} \&\ \ref{spamfig}. In this section we discuss the results for the 
separate sub-samples in the context of the results of our host galaxy study 
(McLure et al. 1999, Dunlop et al. 2000) and previous studies of similar 
objects from the literature. In Sections \ref{hillcomp} \&\ \ref{woldcomp} 
we proceed to perform a more detailed comparison between our results and 
those obtained by Hill \&\ Lilly (1991) and Wold et al. (2000a) at higher 
redshift.
 
\subsection{Do RLQs reside in richer environments than RQQs?}
\label{rqq/rlq}
The clustering results for the two quasar sub-samples shown in 
Tables 2 \&\ \ref{bigtab} 
indicate that there is no significant difference in the richness of the cluster
 environments of the RQQs and RLQs studied here. It can be seen from
Table 2 that the two 
sub-samples display mean and median clustering amplitudes which are
statistically consistent, with the Kolmogorov-Smirnov (KS) test 
returning a significance level of only p=0.54.

The finding that RLQs inhabit environments with spatial clustering amplitudes
 of the order $\sim200\rightarrow300$ Mpc$^{1.77}$ is in good
 agreement with previous 
studies. For example, the large-scale study of Ellingson, Yee \&\ Green (1991)
 found that at $z<0.4$ the average clustering amplitude of RLQs was
 $210\pm70$ Mpc$^{1.77}$. 
Moreover, the clustering study performed by Fisher et al. (1996) using the
 {\sc hst} data for the 20 quasars in the Bahcall et al. (1997) sample 
(all of which are common to the 44 object sample studied in this paper) found 
the 6 RLQs in that sample to have an average clustering amplitude of
 $\simeq256\pm55$ Mpc$^{1.77}$.  
 
In contrast, the finding that the environments of RQQs are indistinguishable
from those of RLQs is not in general agreement with the literature,
with the majority of previous studies concluding that RQQs tend to inhabit
poorer environments. For example, the studies of both Ellingson, Yee
\&\ Green (1991) and Smith, Boyle \&\ Maddox (1995) find the
clustering around RQQs at $z<0.3$ to be perfectly consistent with that
of field galaxies, $B_{gq}/B_{gg}=1.0$, while in contrast even the
median value found here corresponds to $B_{gq}/B_{gg}=3.8$, taking
$B_{gg}=60$. The exact cause of this discrepancy is
difficult to determine due to the large number of factors which
influence the calculation of $B_{gq}$. One possible explanation for the
disagreement between these results and those of Smith et al. is that
$\simeq80\%$ of their objects are fainter than M$_{V}=-23$, which
could explain the detection of environments typical of Seyfert
galaxies (eg. de Robertis \&\ Yee 1998). This issue was looked at by
Smith et al., who found no significant difference in their results
if they divided their objects into high and low luminosity sub-samples,
although this split would still have resulted in $\sim 50\%$ of the
objects in the high-luminosity sub-sample being fainter than M$_{V}=-23$.

It is worth noting at this point that the results presented by Fisher et
al. (1996) of the clustering around the quasars in the Bahcall et
al. (1997) sample do agree with the results presented here. Fisher et
al. found an average value of $B_{gq}=246\pm68$ Mpc$^{1.77}$ for Bahcall's sample
of 14 RQQs, data for all of which are included in the 21 RQQs studied
here. Given that both of our studies are based on {\sc hst} data this
obviously raises the concern that the small field of view
of WFPC2 has led to an overestimate of
the richness of the RQQs environments. However, there are several
reasons for believing that this is not a serious issue. Firstly, as
demonstrated by Fig \ref{angleprof}, there is no real suggestion that
the enhancement in associated galaxies is falling off faster than
$\gamma=1.77$. Secondly, given the good agreement between our results
for the RLQ sub-sample and previous studies, the form of clustering
around the RQQs would have to be substantially different from that
around RLQs in order to account for the difference. Although this is
undoubtedly a possibility, it is not supported by our data, with the
F675W images showing no tendency for the RQQs to have fewer associated
galaxies at radii of $200<r<700$ kpc. It appears therefore that the
similarity in environments of our RQQ and RLQ samples probably results
from the close matching of the samples in the M$_{V}-z$ plane. This
conclusion is supported by the preliminary results of Wold et
al. (2000b), who also find no difference in cluster environment
between their optically matched samples of RQQs and RLQs in the redshift
range $0.5<z<0.8$.

\subsection{Are the environments of RLQs and RGs consistent with unification?}

With the results of our host-galaxy study showing that our RG and RLQ
samples are virtually identical in terms of scalelength, luminosity,
and $R-K$ colour, it was expected that their cluster environments
would also be indistinguishable. The results presented in Table
\ref{bigtab} and Fig \ref{hist} broadly confirm this expectation,
although it is clear that the cluster environments of the two samples 
are not as similar as suggested by the optical properties of their
host galaxies. The application of the KS test shows the two
distributions to be distinguishable only at the $1\sigma$ level ($p=0.25$),
a difference which is clearly due to the three RGs with
$B_{gq}>1000$ Mpc$^{1.77}$. Considering the large errors associated with the
$B_{gq}$ calculation, and the small number statistics, perhaps the
strongest statement that can be made is that these results present no
problem to the orientation-based unification of RGs and RLQs due to
 differences in cluster environment. The possibility of a correlation
between environment and radio power is investigated in Section \ref{black}
 
\begin{table}
\begin{center}
\begin{tabular}{cccccr}
\hline
\hline
 Source & $z$ & $V$ & $M_{V}$ & P$_{5GHz}$ & $B_{gq}\pm\Delta B_{gq}$  \\
\hline
{\bf RG}  &       &       &          \\
 0230$-$027& 0.239 & 19.2  & $-$20.8&24.84&308$\pm$348 \\
 0307$+$169& 0.256 & 18.8  & $-$21.5&25.52&1067 $\pm$498  \\
 0345$+$337& 0.244 & 19.0  & $-$21.9&25.45&334$\pm$364  \\
 0917$+$459& 0.174 & 17.2  & $-$22.0&25.69&1070$\pm$549  \\
 0958$+$291& 0.185 & 17.3  & $-$22.1&25.30&763 $\pm$471  \\
 1215$-$033& 0.184 & 18.9  & $-$20.5&24.00&90 $\pm$265  \\
 1215$+$013& 0.118 & 17.0  & $-$22.3&23.97&299 $\pm$371   \\
 1330$+$022& 0.215 & 18.3  & $-$21.4&25.35&28 $\pm$258  \\
 1342$-$016& 0.167 & 17.8  & $-$21.6&24.35&1593$\pm$658   \\
 2141$+$279& 0.215 & 18.3  & $-$21.4&25.17&202$\pm$313   \\
\hline
{\bf RLQ}  &       &       &          \\
 0137$+$012 & 0.258 & 17.1 & $-$23.9&25.26&582 $\pm$413 \\
 0736$+$017 & 0.191 & 16.5 & $-$23.8&25.35&67  $\pm$257 \\
 1004$+$130 & 0.240 & 15.2 & $-$25.7&24.94&$-$53 $\pm$240  \\
 1020$-$103 & 0.197 & 16.1 & $-$24.2&24.73&266 $\pm$331 \\
 1217$+$023 & 0.240 & 16.5 & $-$24.3&24.92&304 $\pm$347 \\
 1226$+$023 & 0.158 & 12.9 & $-$27.1&26.47&165 $\pm$295     \\
 1302$-$102 & 0.286 & 15.2 & $-$26.1&25.28&320 $\pm$363      \\
 1545$+$210 & 0.266 & 16.7 & $-$24.4&25.26&383 $\pm$371      \\
 2135$-$147 & 0.200 & 15.5 & $-$24.9&25.27&247 $\pm$324 \\
 2141$+$175 & 0.213 & 15.7 & $-$24.8&24.81&208 $\pm$315 \\
 2247$+$140 & 0.237 & 15.3 & $-$23.9&25.31&139 $\pm$302   \\
 2349$-$014 & 0.173 & 15.3 & $-$24.7&24.86&610 $\pm$443 \\
 2355$-$082 & 0.210 & 17.5 & $-$23.0&24.50&235  $\pm$326 \\
\hline
{\bf RQQ}  &       &       &        \\
 0052$+$251 & 0.154 & 15.9 & $-$23.9&21.55& 700$\pm$478     \\
 0054$+$144 & 0.171 & 15.7 & $-$24.3&21.87& $-$115 $\pm$138    \\
 0157$+$001 & 0.164 & 15.7 & $-$24.2&22.87&138 $\pm$280     \\
 0204$+$292 & 0.109 & 16.0 & $-$23.0&     & 675 $\pm$524      \\
 0205$+$024 & 0.155 & 15.4 & $-$24.5&     & 173 $\pm$299     \\
 0244$+$194 & 0.176 & 16.7 & $-$23.4&21.30&$-$123 $\pm$140     \\
 0257$+$024 & 0.115 & 16.1 & $-$23.0&22.19 &$-$53 $\pm$112      \\
 0316$-$346 & 0.265 & 15.1 & $-$26.0&      &227 $\pm$336      \\
 0923$+$201 & 0.190 & 15.8 & $-$24.4&21.26&930 $\pm$501      \\
 0953$+$414 & 0.239 & 15.6 & $-$25.3&21.71&1359 $\pm$553       \\
 1012$+$008 & 0.185 & 15.9 & $-$24.3&22.00 &737 $\pm$459      \\
 1029$-$140 & 0.086 & 13.9 & $-$24.7&      &448$\pm$489      \\
 1116$+$215 & 0.177 & 14.7 & $-$25.5&      &598 $\pm$438        \\
 1202$+$281 & 0.165 & 15.6 & $-$24.5&      &955 $\pm$547        \\
 1307$+$085 & 0.155 & 15.1 & $-$24.8&      &209 $\pm$327       \\
 1309$+$355 & 0.184 & 15.6 & $-$24.7&      &$-$134 $\pm$143      \\
 1402$+$261 & 0.164 & 15.5 & $-$24.5&      &$-$106 $\pm$135        \\
 1444$+$407 & 0.267 & 15.7 & $-$25.4&      &$-$107 $\pm$241        \\
 1635$+$119 & 0.146 & 16.5 & $-$23.1&23.02 &182 $\pm$300         \\
 2215$-$037 & 0.241 & 17.2 & $-$23.7&21.43&$-$52 $\pm$241        \\
 2344$+$184 & 0.138 & 15.9 & $-$23.6&21.11&209 $\pm$314        \\
\hline
\end{tabular}
\end{center}
\caption{Sample details and spatial clustering results. Column 5
details the 5GHz radio luminosity (where known) in units of
WHz$^{-1}$sr$^{-1}$. The errors have been calculated using the
conservative prescription of Yee \&\ L\'{o}pez-Cruz (1999): $\frac{\Delta
B_{gq}}{B_{gq}}=\frac{\left[\left(N_{t}-N_{b}\right)+1.3^{2}N_{b}\right]
^{1/2}}{N_{t}-N{b}}$ 
}
\label{bigtab}
\end{table}
 
\subsection{Abell classification}
\label{abellclass}

In this section we attempt to provide a transformation between the
spatial clustering amplitudes and the traditional Abell cluster
classification (Abell 1958, Abell, Corwin \&\ Olowin
1989). Unfortunately, due to numerous sources of systematic error
associated with Abell counts, the correlation between Abell class and
$B_{gq}$ has a large scatter, and a consensus on the appropriate
calibration has not been reached in the literature. In their study of
47 $z<0.2$ Abell clusters Yee \&\ L\'{o}pez-Cruz (1999) proposed a
linear scheme whereby adjacent Abell classes are separated by $\Delta
B_{gq}=400$ Mpc$^{1.77}$, with a normalization such that Abell class 0 clusters
have a spatial clustering amplitude of $B_{gq}=600$ Mpc$^{1.77}$. Here
we have chosen to classify our cluster measurements using a scheme
which is identical to that proposed by Yee \&\ L\'{o}pez-Cruz save for
a re-calibration such that Abell class 0 clusters correspond to
$B_{gq}=300$ Mpc$^{1.77}$ (Table \ref{abell}). 
 
The main reason for this re-calibration is that, unlike the clusters of
class $\ge1$, the Abell 0 clusters studied by Yee \&\ L\'{o}pez-Cruz
were not selected randomly, but specifically chosen as being
rich. Consequently, it is almost certainly the case that a figure of
$B_{gq}=600$ Mpc$^{1.77}$ is not representative of Abell 0 clusters as
a whole.  Moreover, spatial clustering amplitudes of 
$B_{gq}=300$ \&\ $700$ Mpc$^{1.77}$ for Abell classes 0 \&\ 1 are much 
more representative
of the findings of several previous studies (eg. Longair
\&\ Seldner 1979, Prestage \&\ Peacock 1988, Andersen \&\ Owen
1994). Evidence that the calibration proposed here is also reasonable
for the very richest clusters comes from the fact that the four
richest clusters studied by Yee \&\ L\'{o}pez-Cruz had a mean value of
$B_{gq}=2225$ Mpc$^{1.77}$, here corresponding to Abell class 4/5,
with no clusters found 
with $B_{gq}>2300$ Mpc$^{1.77}$ It can be seen that using our calibration of the
$B_{gq}-$Abell transformation the results presented in this
section imply that the average clustering around our AGN corresponds
to Abell class $\simeq0$, with the large scatter ranging from several
objects consistent with no galaxy enhancement, to several clusters of
Abell class 2/3.

\begin{table}
\begin{center}
\begin{tabular}{ccccccc}
\hline
 Abell Class &0&1&2&3&4&5\\
\hline
 $B_{gq}$&300&700&1100&1500&1900&2300\\
\hline
\end{tabular}
\caption{The proposed relation between spatial clustering amplitude
  and Abell classification. This scheme is identical to that 
 proposed by Yee \&\ L\'{o}pez-Cruz (1999) but with Abell Class 0
 clusters normalized to $B_{gq}=300$ instead of 700 Mpc$^{1.77}$ (see text)}
\label{abell}
\end{center}
\end{table}
 
\section{Do the environments of radio-loud agn change at z=0.5?}
\label{change}
The clustering results for the radio galaxy and radio-loud quasars
sub-samples presented in the previous section offer an opportunity to
re-examine the evidence for an epoch-dependent change in cluster
richness at $z\sim 0.5$. In this section we investigate this issue by
comparing our results with those published for radio galaxies by 
Hill \&\ Lilly (1991) and the recent work on radio-loud quasars by
Wold et al. (2000a).

\subsection{Radio galaxies}
\label{hillcomp}
\begin{figure*}
\centerline{\epsfig{file=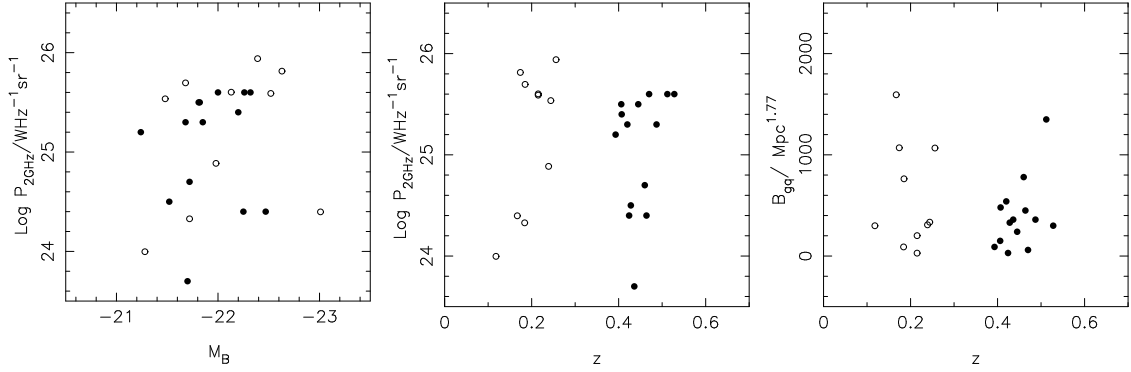,width=15.0cm,angle=0,clip=}}
\caption{The left-hand and middle panels show the matching of our RG
sample (open circles) with the sub-sample of the objects studied by
Hill \&\ Lilly (1991) discussed in the text (filled circles). The right-hand panel
shows a comparison between their respective spatial clustering amplitudes. }
\label{hill_comb}
\end{figure*}

In their study of the environments of radio galaxies Hill \&\ Lilly
(1991) studied four groups of $\simeq 10$ objects at $z\simeq0.5$,
each of
which were equally spaced in radio power from Log
P$_{2GHz}=23\rightarrow27$ WHz$^{-1}$. To compare their results with
those for our radio galaxy sample we selected objects from their two
medium radio luminosity groups, subject to the redshift constraint
$0.40<z<0.53$, which
was chosen to match the range in redshift displayed by our radio
galaxies ($0.12<z<0.26$). This selection procedure produced 16
objects, two of which were quasars and subsequently rejected to leave a final
sample of 14 objects. The P$_{2GHz}-$M$_{B}$ and P$_{2GHz}-z$
distributions of the Hill \&\ Lilly sub-sample and our radio galaxy
sample are shown in the left and middle panels of Fig
\ref{hill_comb}. The two samples are clearly well matched, with the
two-dimensional KS test (2DKS) showing their respective P$_{2GHz}-$M$_{B}$
distributions to be statistically indistinguishable ($p=0.2$). 

To facilitate a comparison between the cluster richness determinations
of Hill \& Lilly and our own results it was necessary to transform
between $B_{gq}$ and their $N_{0.5}$ values. The value of $N_{0.5}$ is
the background subtracted number of galaxies within a radius of 0.5 Mpc,
in the magnitude range $m_{g}\rightarrow m_{g}+3$, where $m_{g}$ is the apparent
magnitude of the radio galaxy. As previously mentioned in Section
\ref{spam}, this magnitude range should be well matched to our choice
of m$_{\star}-1\rightarrow$m$_{\star}+2$, and to make the
transformation we have used Hill \&\ Lilly's own determination of
$B_{gq}=30 N_{0.5}$. The resulting $B_{gq}-z$ distributions for the two
samples are shown in the right-hand panel of Fig \ref{hill_comb}. It
can be seen that the two distributions are very similar, an impression
confirmed by a KS test ($p=0.77$), with no indication that the members
of the low-$z$
sample inhabit poorer environments, or that they display a smaller
range in environment richness. Although we are dealing with small
samples, this result demonstrates that, at least for these two well
matched samples, there appears little evidence for a epoch-dependent
change in {\sc frii} radio galaxy environments at $z\simeq0.5$. In
light of this, we
now move on to look for a change in the environments of radio-loud 
quasars with redshift.

\subsection{Radio-loud quasars}
\label{woldcomp}
\begin{figure*}
\centerline{\epsfig{file=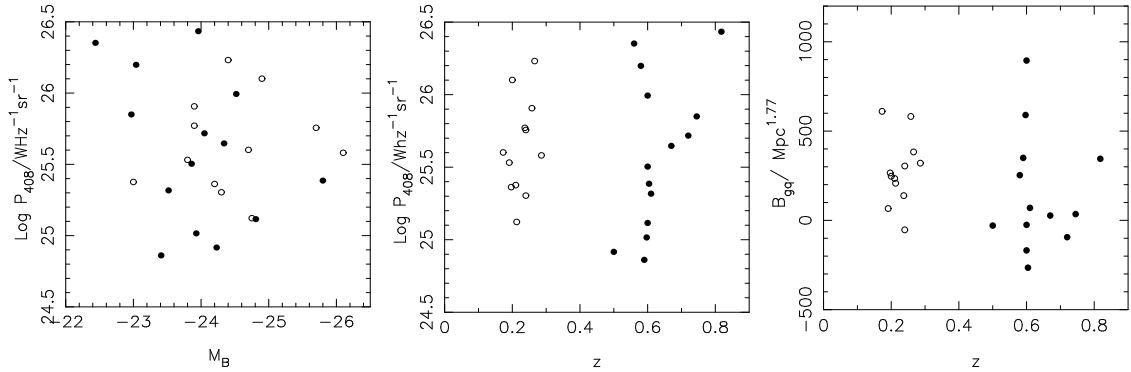,width=15.0cm,angle=0,clip=}}
\caption{The left-hand and middle panels show the matching between our
RLQ sample (open circles) and the sub-sample of the objects studied by 
Wold et al. (2000a) discussed in the text (filled circles). The right-hand
panel shows a comparison of their respective spatial clustering amplitudes.}
\label{wold_comb}
\end{figure*}

The recently published study of 21 RLQs in the redshift
interval $0.5<z<0.8$ by Wold et al. (2000a) provides a good opportunity
for comparison with the results from our $z\simeq0.2$ RLQ sample. Wold
et al. also classify the environments of their quasars via the spatial
clustering amplitude and their results can therefore be directly
compared to our own. Taken as a whole the results for their sample are
in excellent agreement with our own, with their mean figure of
$B_{gq}=265\pm 65$ Mpc$^{1.77}$ being almost identical to our mean of
$B_{gq}=267\pm 51$ Mpc$^{1.77}$. 

To disentangle the effects of redshift and radio luminosity we have
excluded the 7 most radio luminous of the Wold et al. sample (Log
P$_{408}>26.5$ WHz$^{-1}$sr$^{-1}$) to leave a sub-sample of 14
objects which are well matched to our 12-object RLQ sample. The
distribution of the two samples on the P$_{408}-$M$_{B}$ and
P$_{408}-z$ planes is shown in the left-hand and middle panels of Fig
\ref{wold_comb}. The two samples can be seen to be well matched, with
the 2DKS test returning a probability of $p=0.51$ that the two samples
are drawn from the same P$_{408}-$M$_{B}$ distribution. From the
comparison of the spatial clustering amplitudes of the two samples
shown in the right-hand panel of Fig \ref{wold_comb}, there is again
no suggestion that the high-$z$ sample displays systematically higher
values of $B_{gq}$, although it is noticeable that the scatter in
$B_{gq}$ is somewhat higher in the high-$z$ sample. This is in fact
the conclusion arrived at by Wold et al., who also found no
evidence for an epoch-dependent change in RLQ environments when
comparing their results with those of Ellingson, Yee \&\ Green (1991)
at low redshift. 

\section{Evidence for a link between AGN properties and host cluster 
environment}
\label{black}
\begin{figure*}
\centerline{\epsfig{file=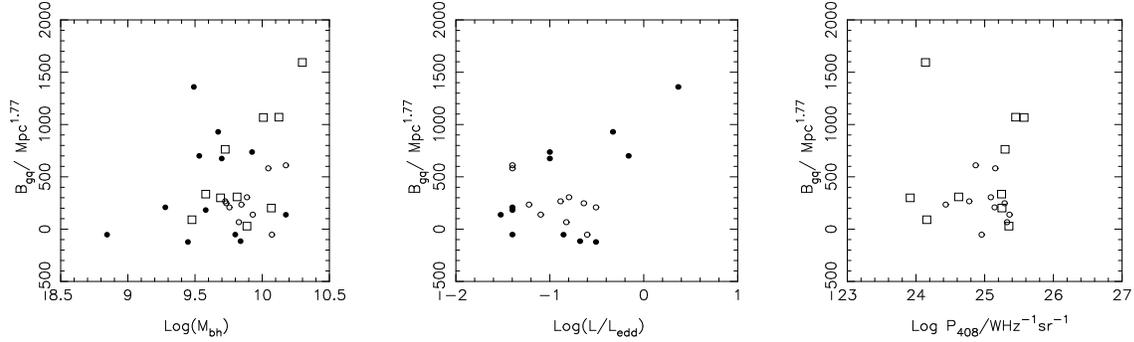,width=15.0cm,angle=0,clip=}}
\caption{Plots of the $B_{gq}$ versus black-hole mass (left), quasar
nuclear luminosity as a fraction of the Eddington limit (middle) and
radio luminosity at 408MHz (right). In all three plots RQQs are represented by
filled circles, RLQs are represented by open circles and RGs are
represented by open squares.}
\label{spamcor}
\end{figure*}

In this section we explore whether there is any evidence in our data for a
link between AGN properties and the large-scale environments of their
hosts. This issue is of interest because it
offers the potential to determine whether the radio and optical
luminosity of powerful AGN are primarily influenced by large-scale
environmental factors such as density of the ICM and galaxy
interactions, or alternatively, whether the
physical properties of the active nucleus itself, such as black-hole
mass and accretion rate, are the dominant influence. 

The three panels
shown in Fig \ref{spamcor} display the relationship between $B_{gq}$
and respectively, estimated central black-hole mass, quasar accretion
efficiency and radio luminosity. All three panels feature only those
objects from the sample imaged in the McLure et al. (1999), and Dunlop
et al. (2000) host-galaxy study in order to produce a homogeneous
data-set. The objects from the Bahcall et al. (1997) sample have been
excluded on the basis that, due to the saturated nature of their data,
reliable nuclear luminosities are unavailable.

In the left-hand panel of Fig \ref{spamcor} we have plotted $B_{gq}$
against black-hole mass, as calculated by a straight application of the
galaxy bulge mass-blackhole mass relation published by Magorrian et
al. (1998) to the host galaxy modelling results presented in Dunlop et
al. (2000). A general trend for black-hole mass to increase with
spatial clustering amplitude can be
seen, although the correlation is not statistically significant, with
a Spearman rank test returning a probability of $p=0.38$ of no
correlation. The clearest signal comes from the ten radio galaxies, 
although even here Kendall's tau test still returns only a
marginally significant result; $p=0.09$. However, the lack of a clear 
correlation between $B_{gq}$ and black-hole mass is perhaps not entirely 
unexpected. 
Even if the basic form of the cluster luminosity function does not change
significantly with cluster richness, there will be a corresponding
increase in the number of luminous,
bulge-dominated galaxies which are capable of hosting powerful AGN
as cluster richness increases. Consequently, in rich clusters,
it is not only the central brightest-cluster galaxy which is a potential AGN
host, but also the next few lower-ranked cluster galaxies. This then
will inevitably introduce significant horizontal scatter into the
$B_{gq}$-blackhole mass relation, regardless of the 0.5 dex scatter in the
Magorrian relation itself.

The middle panel of Fig \ref{spamcor} shows the relationship between
$B_{gq}$ and the nuclear luminosity of the quasars, expressed as a
fraction of their Eddington limit, under the assumption of a central
black-hole mass as predicted by the Magorrian relation. As with the
previous plot, a trend can be seen for the Eddington fraction to
increase with cluster richness, although again this is not
statistically significant, with a rank test returning a probability of
$p=0.5$ that no correlation is present. However, it is perhaps
interesting to note that the three RQQs which are radiating closest to
their predicted Eddington limit are also ranked first, second and
fourth in terms of their cluster richness. 

Finally, the right-hand panel of Fig \ref{spamcor} examines whether 
there is a correlation between the radio luminosity of the
radio-loud AGN in our sample, and the density of the enviroment into
which their radio jets are expanding. As with the other two plots in
Fig \ref{spamcor} a positive correlation is again suggested to the eye, but
is once more found not to be significant, with a rank
test returning a probability of $p=0.47$ of no correlation, even after
the removal of the obvious outlying radio galaxy. It is also
noteworthy that the apparent positive trend between radio luminosity
and $B_{gq}$ is entirely defined by the ten radio galaxies, with the
RLQs showing no evidence for a connection between radio luminosity and
cluster richness. Consequently, at least for this sample of objects,
there appears no real evidence that the density of a radio-loud AGN's
cluster environment plays the dominant role in determining its radio
luminosity. However, it is also clear that the limited dynamic range of
this sample, combined with the inherently large uncertainties
associated with spatial clustering amplitudes, means that the results
presented here are still perfectly consistent with a weak correlation being
present.

\section{Comparing AGN environments to the APM cluster survey}
\label{apm}
\begin{figure}
\centerline{\epsfig{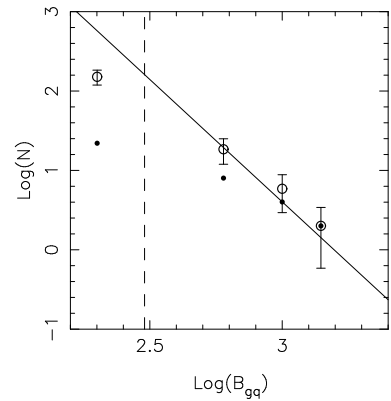}}
\caption{A comparison between our AGN cluster number counts with
those of the APM cluster survey using a bin size of $\Delta
B_{gq}=400$. The solid circles are our uncorrected number counts,
while the open circles are weighted using the inverse of the mean 
$B_{gq}$ in each bin (see text). The solid line is a power-law with 
slope -3.7 which fits the APM number counts  and has arbitrary 
vertical normalization.The vertical dotted line indicates where the 
APM survey becomes incomplete.}
\label{apmfig}
\end{figure}

In Section \ref{change} it was shown that with samples of RQQs and
RLQs which are carefully matched in terms of optical luminosity there
is no evidence that RQQs occupy systematically lower-density cluster
environments. This result therefore makes it of interest to re-examine
the question of whether quasars actually avoid rich clusters, as
sometimes claimed, or if alternatively the apparent lack of 
quasars in rich clusters is simply a reflection of the fact that rich
clusters are relatively scarce.

In order to address this question we have compared our cluster results
with cluster number counts from the APM survey, which are well modelled
by a power-law with slope $\simeq -3.7$ (Dalton et al. 1997). To allow
a fair comparison of the relative numbers of clusters it is necessary
to correct for the fact that
the number of potential AGN host galaxies increases with cluster
richness. To allow for this effect we make the assumption that the
galaxy population of all our clusters can be described by a Schechter
function with constant characteristic magnitude and faint-end slope,
and a normalization simply proportional to $B_{gq}$. Consequently, in
Fig \ref{apmfig} we have simply weighted the relative number counts of
clusters by the inverse of the mean $B_{gq}$ in each bin. As can be
seen in Fig \ref{apmfig} this simple correction naturally produces
number counts which are perfectly consistent with the power-law found
by the APM survey. 

A further interesting result arises if one considers the space
densities of clusters from the APM survey as calculated by 
Croft et al. (1997). With the calibration that in the scheme used in
the analysis of the APM survey, a cluster of richness R=40 corresponds
to roughly Abell class 0 (ie. $B_{gq}\sim 300$ Mpc$^{1.77}$), it is
possible to extrapolate the
figures provided by Croft et al. to estimate that the space density of all
clusters richer than R$\ge40$ is $\simeq 10^{-5}$ Mpc$^{-3}$. If one
allows for the fact that up to 50\% of quasars could be obscured in
the optical-UV by a dusty torus, this abundance of clusters is in
reasonable agreement with the peak space density of quasars at
$z\sim2.5$ (Warren, Hewett \&\ Osmer 1994).  

This line of argument leads to an interesting conclusion concerning
the fraction of massive cluster galaxies which were active at
$z\sim2.5$. In their photometric study of 83 Abell clusters Schneider,
Gunn \&\ Hoessel (1983) determined the absolute magnitudes of the
three brightest galaxies in each cluster. When converted to our
cosmology, the mean magnitude of the first ranked galaxy 
in clusters of Abell class 0 corresponds to  $M_{R}\simeq
-23.1\pm0.5$, where a
colour of $r-R=0.35$ has been assumed (Fukugita et al. 1995). This
is in excellent agreement with the figure of $M_{R}\sim-23$ which was found 
to be the necessary bulge luminosity for a galaxy to host a powerful 
($M_{V}<-23$) quasar by Dunlop et al. (2000). Furthermore, because the
mean luminosity found by Schneider, Gunn \&\ Hoessel for the second
ranked galaxy in Abell class 0 clusters was approximately one magnitude
fainter than that of the first ranked galaxies, we can conclude that
the average number of massive elliptical galaxies in a Abell class 0
cluster which are capable of hosting a powerful quasar is close to
unity. Consequently, because of the steep decline in the number of
clusters with increasing richness (Fig \ref{apmfig}), the close
agreement between the peak space density of quasars and that of clusters of
Abell class $\ge0$ leads to the conclusion that virtually all massive
cluster ellipticals were active at $z\sim2.5$. This conclusion is
unaffected by the fact that the very richest clusters may have
contained 4 or 5 active galaxies at this epoch because the rarity of
such clusters means that they have a negligible affect on quasar numbers. 

\section{Conclusions}
The cluster environments of a matched sample of 44 powerful radio-loud and 
radio-quiet AGN at $z\sim0.2$ have been analysed using the spatial
clustering amplitude method. The three main conclusions arising from
this study can be summarized as follows:

Firstly, we find no evidence to suggest that RQQs inhabit poorer cluster
environments than RLQs or RGs. All three classes of AGN are
located in environments which, on average, are comparable to Abell
Class 0, although there is a large amount of scatter. 

Secondly, by comparing our results with those for samples from Hill
\&\ Lilly (1991) and Wold et al. (2000a) which are matched in terms of 
redshift and radio luminosity, we have re-examined the evidence for an
epoch-dependent change in the environments of RLQs and RGs. We find no
evidence that the environments of RGs and RLQs become significantly
richer at $z\simeq0.5$ as has been reported by Hill \&\ Lilly (1991) 
and Ellingson, Yee \&\ Green (1991) respectively.

Finally, via comparison with the APM cluster survey we conclude that
the distribution of AGN cluster environments is consistent with having
being drawn at random from the general cluster
distribution. Furthermore, because the cluster population is dominated
by clusters of Abell class $\simeq0$ which, on average, only contain
one galaxy capable of hosting a powerful quasar, we argue that the
close agreement between the space density of clusters and the peak
space density of quasars at $z\sim2.5$ suggests that practically all
massive cluster ellipticals were active at this epoch. In this
scenario, the apparent lack of powerful AGN in rich clusters at the
present day simply reflects the comparative scarcity of 
high density environments.

\section{Acknowledgements}
 Based on observations with the NASA/ESA Hubble Space Telescope, 
 obtained
 at the Space Telescope Science Institute, which is operated by the
 Association of Universities for Research in Astronomy, Inc. under NASA
 contract No. NAS5-26555.
 This research has made use of the NASA/IPAC Extragalactic Database (NED)
 which is operated by the Jet Propulsion Laboratory, California Institute
 of Technology, under contract with the National Aeronautics and Space
 Administration. The authors thank Philip Best and Lance Miller
 for their useful comments. RJM acknowledges a PPARC PDF.
\section{References}
\noindent
Abell G., 1958, ApJS, 3, 211\\
Abell G., Corwin H.G., Olowin R.P., 1989, ApJS, 70, 1\\
Andersen V., Owen F.N., 1994, AJ, 108, 361\\
Bahcall J.N., Kirhakos S., Saxe D.H., Schneider D.P., 1997, ApJ, 479, 642\\
Best P.N., Longair M.S., R\"{o}ttgering H.J.A., 1997, MNRAS, 292, 758\\
Best P.N., Longair M.S., R\"{o}ttgering H.J.A., 1998, MNRAS, 295, 549\\
Boyce P.J., et al., 1998, MNRAS, 298, 121\\
Croft R.A.C., Dalton G.B., Efstathiou G., Sutherland W.J.,  Maddox
S.J. 1997, MNRAS, 291, 305\\
De Robertis M.M., Hayhoe K., Yee H.K.C., 1998, ApJS, 115, 163\\
Dalton G.B., Maddox S.J, Sutherland W.J., Efstathiou G., 1997, MNRAS, 289, 263\\
Dunlop J.S., et al., 1993, MNRAS, 264, 455\\
Dunlop J.S., et al., 2000, MNRAS, in prep\\
Ellingson E., Yee H.K.C., Green R.F., 1991, ApJ, 371, 49\\
Fisher K.B., Bahcall J.N., Kirhakos S., 1996, ApJ, 468, 468\\
Fukugita M., Shimasaku K., Ichikawa T., 1995, PASP 107, 945\\
Groth E.J., Peebles P.J.E., 1977, ApJ, 217, 385\\
Hall P.B., Green R.F., 1998, ApJ, 507, 558\\
Hill G.J., Lilly S.J., 1991, ApJ, 367, 1\\
Holtzman J.A., et al., 1995, PASP, 107, 1065\\
Hooper E.J., Impey C., Foltz C.B., 1997, ApJ, L95\\
Laing R.A., Riley J.M., Longair M.S., 1983, MNRAS, 204, 151\\
Lilly S.J., Prestage R.M., 1987, MNRAS, 225, 531\\
Lilly S.J., et al., 1995, ApJ, 455, 108\\
Longair M.S., Seldner M., 1979, MNRAS, 189, 433\\
Loveday J., Peterson B.A., Efstathiou G., Maddox S.J., 1992, ApJ, 390, L338\\
Magorrian J., et al., 1998,  AJ, 115, 2285\\
McLure R.J., et al., 1999, MNRAS, 308,377\\
Metcalfe N., Shanks T., Fong R., Jones L.R., 1991, MNRAS, 249, 498\\
Prestage R.M., Peacock J.A., 1988, MNRAS, 230, 131\\
Prestage R.M., Peacock J.A., 1989, MNRAS, 236, 959\\
Schade D.J., Boyle B.J., Letawsky M., 2000, MNRAS, 315, 498\\
Schneider D.P., Gunn J.E., Hoessel J.G., 1983, ApJ, 268, 476\\
Smith E.P., O'Dea C.P., Baum S.A., 1995, ApJ, 441, 113\\ 
Smith R.J., Boyle B.J., Maddox S.J., 1995, MNRAS, 277, 270\\
Smith R.J., Boyle B.J., Maddox S.J., 2000, MNRAS, 313, 252\\
Stocke J.T., et al., 1983, ApJ, 273, 458\\
Taylor G.L., et al., 1996, MNRAS, 283, 930\\
Warren S.J., Hewett P.C., Osmer P.S., 1994, ApJ, 421, 412\\
Wold M., Lacy M., Lilje P.B., Serjeant S., 2000a, MNRAS, submitted
(astro-ph/9912070)\\
Wold M., Lacy M., Lilje P.B., Serjeant S., 2000b, (astro-ph/0006063)\\
Wurtz R., Stocke J.T., Ellingson E., Yee H.K.C., 1997, ApJ, 480, 547\\
Yates M.G., Miller L., Peacock J.A., 1989, MNRAS, 240, 129\\ 
Yee H.K.C., Green R.F., 1984, ApJ, 280, 79\\
Yee H.K.C., Green R.F., 1987, ApJ, 319, 28\\
Yee H.K.C., Ellingson E., 1993, ApJ, 411, 43\\
Yee H.K.C., L\'{o}pez-Cruz O., 1999, AJ, 117, 1985\\          
\end{document}